\documentstyle[12pt,psfig]{article}
\newcommand\new{\newcommand}         % shorthand for \newcommand
\newcommand\ren{\renewcommand}       % shorthand for \renewcommand
\ren{\textfraction}{0.01}            % accept a text page with 1% text
\ren{\topfraction}{0.99}             % accept a text page with 99% figure
\ren\d{\mathrm{d}}                   % replace an accent...
\new\hf{\hspace*{\fill}}  % sure horizontal fill
\new\vs{\vspace*{5mm}}    % sure 5mm vertical space
\new\cb{\begin{center}}              
\new\ce{\end{center}}
\new\qb{\begin{quote}}
\new\qe{\end{quote}}
\new\CENTER[1]{  \hf #1 \hf} 
\new\RIGHT[1]{   \hf #1}
\new\CERN{       \cb{\large EUROPEAN ORGANIZATION FOR NUCLEAR RESEARCH}\ce\vs}
\new\PPE[1]{     \hf CERN~/~PPE~#1}
\new\ALEPH[1]{   \hf\makebox[4.2em][l]{ALEPH}  \makebox[3.2em][r]{#1}}
\new\PHYSIC[1]{  \hf\makebox[4.2em][l]{PHYSIC} \makebox[3.2em][r]{#1}}
\new\DRAFT{      \vs\vs\cb{\bf\Huge DRAFT DRAFT DRAFT DRAFT}\ce\vs\vs}
\new\TITLE[1]{   \cb{\large\bf #1}\ce}
\new\AUTHOR[1]{  \vs\cb\large\bf #1 \ce\vs}
\new\ABSTRACT[1]{\cb{\large\bf Abstract}\ce\normalsize\qb #1 \qe\vs\vs}
\new{\be}{\begin{equation}}
\new{\eeq}{\end{equation}}
\new{\beq}[1]{\begin{equation}\label{eq:#1}}
\new{\eq}[1]{\ref{eq:#1}}
\new{\fig}[1]{Fig.~\ref{fig:#1}}
\new{\Fig}[1]{Figure~\ref{fig:#1}}
\new{\tab}[1]{Tab.~\ref{tab:#1}}
\new{\Tab}[1]{Table~\ref{tab:#1}}
\new{\dsty}{\displaystyle}
\new{\mbf}{\boldmath}
\new{\mrm}{\mathrm}
\new{\mm}[1]{{\mbox{\hspace{#1mm}}}}
\new{\nin}{\noindent}
\new{\LambdaDIS}{\mbox {$\Lambda_{\overline{\rm\tiny MS}}^{{\small(4)}}$}}
\new{\as}{\mbox{$\alpha_s$}}
\new{\ee}{\mbox{$e^{+}e^{-}$}}
\new{\bb}{\mbox{$b\overline{b}$}}
\new{\cc}{\mbox{$c\overline{c}$}}
\new{\pp}{\mbox{$p\overline{p}$}}
\new{\AS}{\mbox{$\mbf{\alpha_s}$}}
\new{\aem}{\mbox{$\alpha_{em}$}}
\new{\asmz}{{\mbox{$\alpha_s(M_Z)$}}}
\new{\asmw}{{\mbox{$\alpha_s(M_W)$}}}
\new{\asmu}{{\mbox{$\alpha_s(\mu)$}}}
\new{\asmtau}{\mbox{$\alpha_s(m_{\tau})$}}
\new\Rtau{\mbox{$R_{\,\displaystyle\tau}$}}
\new\Rgamma{\mbox{$R_{\,\gamma}$}}
\new\Rl{\mbox{$R_{l}$}}
\new{\MeVcc}{\mbox{${\mrm{MeV}}/c^2$}}
\new{\GeVcc}{\mbox{${\mrm{GeV}}/c^2$}}
\new{\GeV}{\mbox{\rm~GeV}}
\new\pbp{\mbox{$p\overline{p}$}}
\new\into{\mbox{$\rightarrow$}}
\new\ass{\mbox{$\alpha_s^2$}}
\new\asss{\mbox{$\alpha_s^3$}}
\new\Oas[1]{\mbox{${\cal O}(\alpha_s^{#1})$}}
\new\MSbar{\mbox{$\overline{\mrm{MS}}$}}
\new\expt{\mbox{exp}}
\new\theo{\mbox{theo}}
\new\errsx[2]{${\dsty #1} \pm {\dsty #2}$}
\new\errax[3]{${\dsty #1}\mm{0.2}\pm\mm{0.2}^{#2}_{ #3}$}
\new\errs[2]{${\dsty #1}\,{(\dsty #2)}$}
\new\erra[3]{${\dsty #1}\,(\mm{0.3} ^{#2}_{#3}\mm{0.1})$}
\new\xbib[3]{\bibitem{ref:#1}{#2}} 
\new{\etal}{\mbox{\it et al.}}
\new\NCA[1]{{\em Nuovo Cimento} #1}
\new\NIM[1]{{\em Nucl. Instrum. Methods} #1}
\new\NIMA[1]{{\em Nucl. Instrum. Methods} A #1}
\new\NPB[1]{{\em Nucl. Phys.} B #1}
\new\NPBP[1]{{\em Nucl. Phys.} B (Proc. Suppl.) #1}
\new\PLB[1]{{\em Phys. Lett.} B #1}
\new\PRL[1]{{\em Phys. Rev. Lett.} #1}
\new\PRB[1]{{\em Phys. Rev.} B #1}
\new\PRD[1]{{\em Phys. Rev.} D #1}
\new\PR[1]{{\em Phys. Rev.} #1}
\new\PRP[1]{{\em Phys. Rep.} #1}
\new\ZPC[1]{{\em Z. Phys.} C #1}
\new\ANP[1]{{\em Annals of Physics} #1}
\new\PSC[1]{{\em Physica Scripta} #1}
\new\JMP[1]{{\em J. Math. Phys.} #1}
\new\IJMPA[1]{{\em Int. J. Mod. Phys. A} #1}
\new\SJNP[1]{{\em Sov. J. Nucl. Phys.} #1}
\new{\Coll}{Collaboration}
\new{\hepGlasgow}{ICHEP conference, Glasgow, 20-27 July 1994}
\new{\hepBrussels}{EPS-HEP Conference, Brussels, 27 July - 2 August 1995} 
\new{\hepWarsaw}{ICHEP conference, Warsaw, 25-31 July 1996}
\new{\qcdAachen}{{\em QCD 20 years later}, Aachen, 9-13 June 1992}
\new{\latDallas}{{\em Lattice `93}, Dallas, October 12-16, 1993}
\new{\Mxxvii}{{\em XXVIIth Rencontres de Moriond}, 22-28 March 1992}
\new{\Mxxx}{{\em XXXth Rencontres de Moriond}, 19-26 March 1995}
\new{\Mxxxi}{{\em XXXIth Rencontres de Moriond}, 16-23 March 1996}
\new{\tauArgonne}{{\em Workshop on Tau-Charm Factory}, Argonne, June 1995}
\new{\tauMontreux}{{\em 3rd Workshop on Tau Lepton Physics},
                    Montreux, 19-22 September 1994}
\new\fplace[3]{
\begin{minipage}[h]{#1\textwidth}
  \vspace*{#2}
  \begin{center}
      \mbox{\psfig{file=#3,width=\textwidth}}
  \end{center}
 \end{minipage}}
\new\cplace[4]{
\begin{minipage}[h]{#1\textwidth}
  \vspace*{#2}
  \caption[]{#4}
      \label{#3}
 \end{minipage}}
\begin{document}
\pagestyle{empty}
\setcounter{page}{1}
\vspace*{2cm}
\TITLE{STATUS OF THE STRONG COUPLING CONSTANT}
\AUTHOR{Michael Schmelling}  
\CENTER{Max-Planck-Institut f\"ur Kernphysik} \\
\CENTER{Postfach 103980, D-69029 Heidelberg}
\vspace*{\fill}
\ABSTRACT{
   The current status of measurements of the strong coupling
   constant from different reactions is reviewed. Including
   new results presented at the 1996 ICHEP conference, a global
   average $\asmz=0.118 \pm 0.003$\/ is obtained.}
\vspace*{\fill}
\begin{center}
  Plenary talk given at the                                      \\
  {\em XXVIII International Conference on High Energy Physics}   \\
  Warsaw, July 25--31, 1996.
\end{center}
\vspace*{\fill}
\newpage
\pagestyle{plain}

\section{Introduction}
%=====================
Over the past years significant progress has been made in the
determination of the strong coupling constant \as. Next-to-leading order
(NLO) theoretical predictions are generally available. For some inclusive
quantities also the next-to-next-to-leading (NNLO) orders have been calculated
and estimates of the next higher terms exist. In addition, resummations of 
leading-log (LL) and next-to-leading-logarithmic (NLL) corrections to
all orders have been performed in some cases. Power law corrections are
controllable in the framework of the operator-product-expansion
(OPE) or the resummation of renormalon chains. On the experimental side
information exists from many different reactions over a large range of
both space-like and time-like momentum transfers. Reactions include
neutrino- and lepton-nucleon scattering, proton-(anti)proton collisions,
\ee-annihilation and decays of bound states of heavy quarks. Observables
are total cross section measurements, sum rules, scaling violations, 
branching ratios, global event shape variables and production rates
of hadron jets, i.e. bundles of particles close by in phase space originating 
from isolated hard partons.

\subsection{Theoretical Predictions}
%===================================
The QCD prediction for a cross section $\sigma(Q)$\/ at an energy scale
$Q$\/ can be expressed as sum of perturbative terms $\delta_{pert}$,
which vary logarithmically with energy, and non-perturbative power law
corrections $\delta_{np}$. The perturbative part is of the form 
$$
   \delta_{pert}
 = \as(\mu)A + \ass(\mu)
   \left( B + A\frac{\beta_0}{4\pi} \ln\frac{\mu^2}{Q^2} \right)
 + \ldots
$$
Here $\mu$\/ is an arbitrary renormalization scale used in the
calculation and $\beta_0$\/ the leading order coefficient of the QCD
beta-function. For $\mu \approx Q$\/ the perturbative expansion is well
behaved, allowing for a measurement of the strong coupling constant.
For $|\ln(\mu^2/Q^2)| \gg 1$\/ the effective expansion parameter becomes
$\as(\mu) \ln(\mu^2/Q^2)$, which leads to an unstable theoretical
prediction. The full theory does not depend on $\mu$, but the truncated
perturbative expansion does. Variation of $\mu$\/ thus allows to probe
the sensitivity of the theory to uncalculated higher orders and thereby
to assess theoretical uncertainties.

In the following the nominal scale $\mu$\/ of an \as-measurement is
identified with the physical scale of the respective process. 
For data covering a range of scales the geometric mean of high and low 
end is taken, unless a central scale is specified explicitly by the 
authors. Combined results from different measurements are quoted at the 
geometric mean of the individual scales. The geometric mean for the 
scales is motivated by the fact, that (in leading order) \as\/ is 
inversely proportional to the logarithm of the scale.

The beta function describes the renormalization scale dependence of
the strong coupling,
$$
      \mu \frac{d\as(\mu)}{d\mu}
   = -\frac{\beta_0}{2\pi} \alpha_s^2(\mu)
     -\frac{\beta_1}{4\pi^2} \alpha_s^3(\mu)
     + {\cal O}(\alpha_s^4)
$$ 
with $\beta_0=11-2 n_f/3$\/ and $\beta_1=51-19 n_f/3$\/ for QCD based on an 
SU(3)-colour gauge symmetry. The quantity $n_f$\/ denotes the number of active
quark flavours. Measurements of \as\/ obtained at different energy scales can 
be compared, either by evolving backwards to the point $\Lambda$\/ where \as\/
diverges, or by evolving to a common reference energy, which in recent years 
has become the Z mass. In the evolution~\cite{ref:RUNNING} 
\as\/ is continous at flavour thresholds, which, within the still 
comparatively large uncertainties of the quark masses, can be taken at the 
\MSbar-running mass of the respective quark flavour. As an example 
consider the evolution of an \as-measurement with $n_f=3$\/ from the scale
$\mu=m_{\tau}$\/ to $\mu=M_Z$. It proceeds by first going back to the
charm-quark mass $m_c = 1.3 \pm 0.2$~GeV using the evolution equation for
three flavours, then going up to $m_b = 4.3 \pm 0.2$~GeV with $n_f=4$\/
and finally evolving all the way up to $\mu=M_Z$\/ with $n_f=5$\/ active
flavours. Many slightly different schemes are in use to perform the
evolution numerically. Despite being formally equivalent to NLO the
resulting differences in the value for \asmz\/ can be as large as 
$\delta\asmz=0.001$. To be consistent, for the numbers presented in this 
paper the treatment of flavour thresholds described above has been 
applied together with Runge-Kutta integration of the NNLO-beta function.

\subsection{Combination of Individual Results}
%=============================================
Many measurements of the strong coupling constant are dominated by
theoretical uncertainties. In order to be quantitative, it will be assumed 
that they represent 68\% confidence intervals. Their non-statistical nature 
suggests that they should be interpreted in the Bayesian sense. Note that 
doing this also for the experimental uncertainties would justify to combine 
experimental and theoretical errors in quadrature.

Theoretical errors for different \as-measure\-ments are correlated via
the underlying theory. Such correlations could be quantified by the
derivatives of the predictions with respect to the various sources of
uncertainty, e.g. unknown higher order coefficients. Instead, in many 
cases only total errors are available, leading to a situation where
measurements are known to be correlated with very little information about
the actual size of those correlations. To deal with this sitution, the
following averaging procedure~\cite{ref:MSAV} has been employed.

For a set of measurements $x_i\pm\sigma_i$\/ the average is calculated
as $\overline{x}=\sum x_i w_i$, with weights inversely proportional
to the squares of the errors $\sigma_i$. This is a robust estimate of
a common mean, which is also optimal if the single measurements are
uncorrelated. To determine the error of the average and its $\chi^2$\/
the full covariance matrix is needed. Using only the diagonal terms,
both are underestimated in the presence of positive correlations, i.e.
a $\chi^2$\/ smaller than its expectation value is indicative 
of such terms. In this case a realistic error for the average can be
obtained even if the size of the correlations is not known a priori,
by introducing off-diagonal terms $\rho\sigma_i\sigma_j$\/ in the
covariance matrix and adjusting the effective correlation coefficient 
$\rho$\/ such that $\chi^2$\/ attains its expectation value. Averages
with errors determined according to this method will be referred to
as ``conservative averages'' in the following. Measurements with
asymmetric errors are shifted to the center of the error band and the
error symmetrized prior to the averaging procedure.

\section{Measurements of \AS}
%============================
This section contains an overview over the most important types of
\as-measurements, presenting results from a large variety of reactions
and observables.

\subsection{Inclusive Measurements}
%==================================
Inclusive measurements of \as\/ depend only on one energy scale
characterizing the process and are best understood theoretically.
The perturbative prediction is known to \Oas{3}, non-perturbative
effects can be treated in the framework of the Operator Product
Expansion~\cite{ref:SVZ}(OPE).

\subsection*{The Ratios \Rgamma\/ and \Rl}
%=========================================
The quantities \Rgamma\/ and \Rl\/ are defined by the ratios of
the hadronic to the leptonic branching fractions of a virtual
photon or the Z-boson. In both cases the hadronic system is formed
from the electroweak (EW) coupling of a vector boson (Photon or Z) to 
a primary quark-antiquark pair. The sensitivity to the strong coupling  
comes from gluon radiation off the primary quarks. This radiation
opens up new final states for the hadronic system, which increase the
hadronic width with respect to the purely electroweak expectation.
The QCD correction is dominated by the perturbative terms. The leading
non-perturbative effects scale with ${\cal O}(1/Q^4)$\/ and thus are
negligible for \Rgamma\/ and \Rl. Mass effects, in particular from the
large top-bottom mass splitting, have been calculated~\cite{ref:RMB}
to \Oas{2}.

The first analyses~\cite{ref:RGAMMA1} of \Rgamma\/ were still based 
on an erroneous third order correction~\cite{ref:RK3BAD}. A recent
reanalysis~\cite{ref:RGAMMA2} using the correct third order
coefficient~\cite{ref:RK3OK} and data from centre-of-mass energies 
between 5~GeV and 65~GeV yields $\asmu=0.175 \pm 0.023$\/ for a central 
scale $\mu=18$~GeV, which corresponds to $\asmz=0.128 \pm^{0.012}_{0.013}$. 

A simple parametrization~\cite{ref:RL} for \Rl\/ is available to
third order in \asmz. Using the combined result from all four
LEP-experiments~\cite{ref:RL96} $R_l=20.778 \pm 0.029$, a top-quark mass 
$m_t=180$~\GeVcc\/ and a Higgs-mass $M_H=300$~\GeVcc\/ one finds
$\asmz=0.124 \pm 0.005$. The statistical error is 
$\delta\asmz_{stat}=0.004$, twice as large as the QCD theoretical
uncertainty. 

\begin{figure}[htb]
\begin{center}
   \fplace{0.60}{0.0cm}{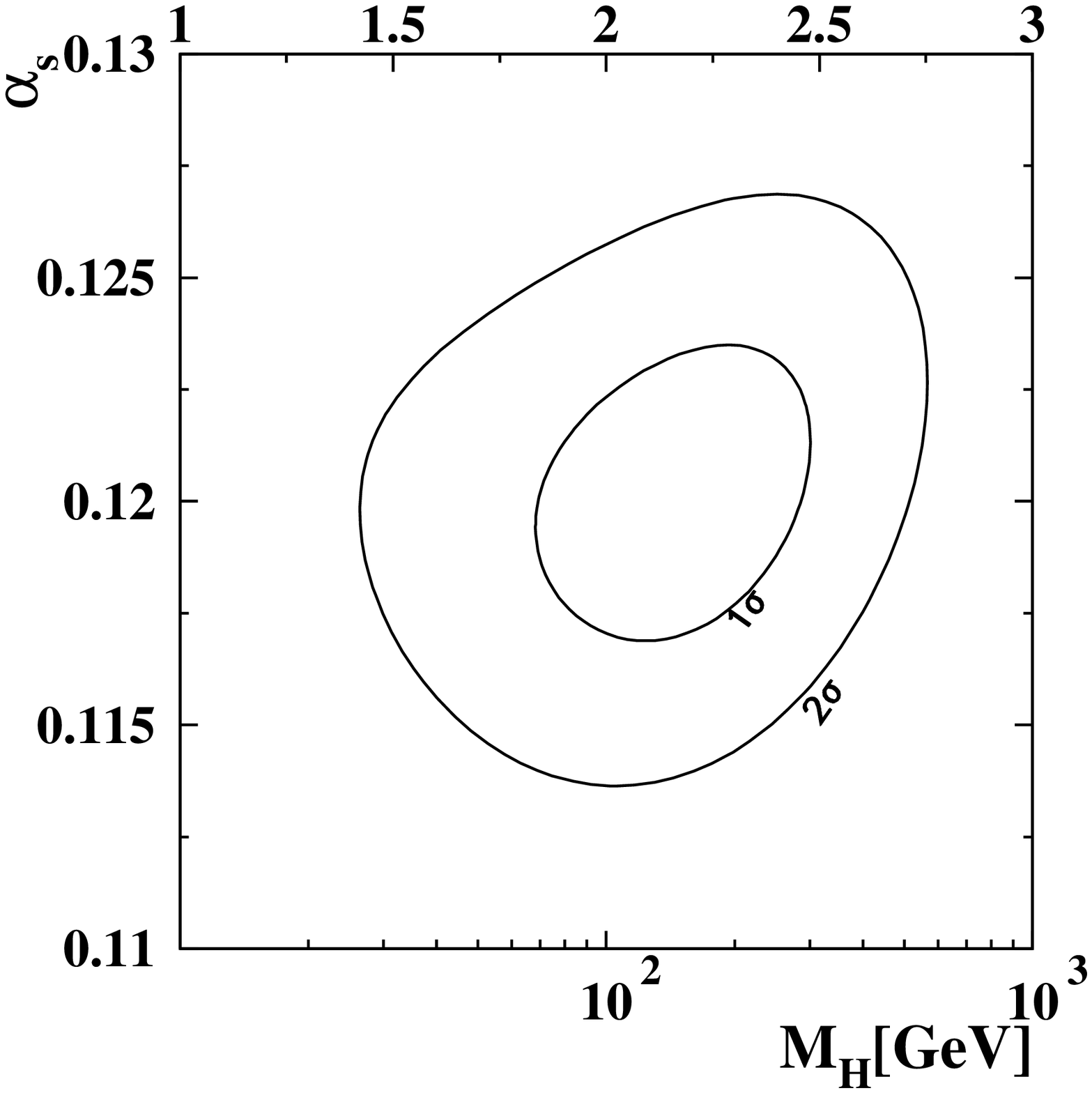}
   \cplace{0.35}{3.9cm}{fig:ASEW}
          {\sf Combined measurement of \as\/ and the Higgs mass 
            $M_H$\/ from QCD radiative corrections to the standard model.}
\end{center}
\end{figure}

In addition to \Rl\/ also the hadronic peak cross section of the Z and
its width are sensitive to the strong coupling. Combining this information
with all available electroweak data from LEP, SLC, collider measurements
and Deep Inelastic Scattering in a global standard model fit, allows the 
simultaneous determination of \asmz\/ and the Higgs mass. The result is 
shown in \fig{ASEW}. One finds~\cite{ref:RBDL} a Higgs mass 
$M_H=149$~\GeVcc\/ and $\asmz=0.1202 \pm 0.0033$. The standard model 
constraint thus not only gives a very precise measurement of \asmz, but 
also one perfectly consistent with other precision measurements.

\subsection*{Measurement of \as\/ from \Rtau}
%============================================
An \as-measurement can also be obtained from $\Rtau=B_{\mrm{hadr}}/B_e$,
the ratio of the hadronic to the electronic branching ratio of the
tau lepton. Here QCD radiative corrections affect the hadronic final
state from a W-decay. In contrast to \as-determinations from \Rgamma\/
or \Rl\/ the mass of the hadronic system is not fixed in tau decays.
This makes the quantity \Rtau\/ double inclusive, i.e. integrated over
all hadronic final states at a given mass and integrated over all masses
between $M_{\pi}$\/ and $m_{\tau}$.

The low energy scale requires a good understanding of the non-perturbative
contributions, which in the framework of the OPE are proportional to
vacuum expectation values (condensates) of the QCD fields. These 
condensates can be determined from independent phenomenological 
analyses~\cite{ref:RTAU1ST}, or together with \as\/ from the higher 
moments of the mass spectrum of hadronic $\tau$-decays.\cite{ref:RTAU2ND}
It turns out that the non-perturbative corrections to \Rtau\/ are surprisingly
small, $\delta_{np}=-1.5 \pm 0.4 \%$, and~\cite{ref:RTAUMN} that 
perturbative QCD is applicable down to mass scales below 1~GeV. 
As a consequence an \as\/ measurement from \Rtau\/ is potentially 
very accurate.

Assuming the validity of the completeness relation for the tau branching
ratios into hadrons, electrons and muons, $B_{\mrm{hadr}}+B_e+B_{\mu}=1$,
and lepton universality, the ratio \Rtau\/ can be expressed as function
of $B_e$\/ alone. Note that the larger mass of the muon leads to 
$B_{\mu}/{B_e} = 0.9726$. The branching ratio $B_e$\/ can be determined 
by direct measurements or, again assuming lepton universality, by comparing 
mass and lifetime of the tau lepton and the muon. From recent 
measurements~\cite{ref:RTAUPAR} one obtains $\Rtau=3.642 \pm 0.010$. A value
for \asmtau\/ with a rather conservative error based on this number is
quoted~\cite{ref:RTAUMN} as $\asmtau = 0.33 \pm 0.03$. This error is
almost entirely due to uncertainties in the perturbative prediction
$\delta_{pert}$, taken to be half the difference between
Le~Diberder-Pich resummation~\cite{ref:RTAURS1} and resummation of
renormalon chains.\cite{ref:RTAURS2} Other studies~\cite{ref:RTAUNEW}
suggest that it could be significantly smaller.

A combined analysis of \Rtau\/ and the leading moments of the hadronic
mass spectrum has been performed by the ALEPH~\cite{ref:RTAUALEPH} and
CLEO~\cite{ref:RTAUCLEO} collaborations. A three-sigma discrepancy
between the two was due to the old 1994 PDG-value~\cite{ref:PDG94} for
$B_e$\/ used in the CLEO analysis. The moments extracted from the hadronic
mass spectrum are in good agreement. With an upda\-ted leptonic branching
ratio~\cite{ref:RTAUBENEW} the CLEO result $\asmtau=0.339 \pm 0.024$\/
becomes compatible with the ALEPH number $\asmtau=0.353 \pm 0.022$.
The conservative average of both \as-measurements is
$\asmtau=0.347\pm 0.022 \pm 0.03$, where the second error~\cite{ref:RTAUMN}
has been added to account for the fact that both analyses are based on 
Le Diberder-Pich resummation. Evolving up to the Z-mass and symmetrizing 
the larger one of the slightly asymmetric errors, one finally obtains
$\asmz = 0.122\pm 0.005$.

\subsection*{\as\/ from the Gross-Llewellyn-Smith Sum Rule}
%==========================================================
Determinations of the strong coupling constant based on sum rules are
fully inclusive measurements at a very low $Q^2$-scale. In the
Quark-Parton-Model (QPM) the Gross-Llewellyn-Smith sum rule (GLSR)
counts the number of valence quarks in the nucleon. The perturbative
QCD-correction~\cite{ref:GLS1} is known to \Oas{3}. A measurement of
the strong coupling constant taking into account also non-perturbative
(``higher twist'') terms~\cite{ref:GLS2} has been performed by the
NuTeV-Collaboration~\cite{ref:GLS3} at a scale $\mu=1.73$~GeV, based on the 
old CCFR data.\cite{ref:GLS4} The result $\asmu=0.260^{+0.041}_{-0.046}$\/ 
corresponds to $\asmz=0.110^{+0.006}_{-0.009}$. As the input data have been
re-calibrated recently, this number can be expected to be updated, too.

\subsection*{The Bjorken Sum Rule}
%=================================
Also for the Bjorken sum rule (BjSR) the perturbative QCD-correction to
the QPM~\cite{ref:BJSR1} is known to \Oas{3} together with an estimate
of the size of the non-perturbative effects.\cite{ref:BJSR2} The BjSR
is defined for charged-current neutrino-nucleon scattering or polarized
lepton nucleon scattering. Measurements of the strong coupling constant
so far only exist based on the spin-dependent structure functions of
the nucleon. A first analysis~\cite{ref:BJSR3} had comparable experimental 
and theoretical errors, a recent update~\cite{ref:BJSR4} finds 
$\asmu=0.320^{+0.031}_{-0.053}(\expt)\pm 0.016(\theo)$\/ for a 
scale $\mu=1.73$~GeV. Evolved to the Z-mass one obtains  
$\asmz=0.118^{+0.005}_{-0.007}$\/ with an error dominated by the 
experimental uncertainties.

The small size $\delta\asmu=0.016$\/ of the theoretical uncertainties 
results from the stability of the analysis with respect to various ways of 
estimating missing higher order terms by means of Pad\'e approximants (PA).
The PA $[N/M]$\/ of a function is the ratio of two polynomials of order
$N$\/ and $M$, which to order $N+M$\/ has the same
Taylor-expansion as the original function. Pad\'e approximants thus
offer a systematic way to guess how a perturbative expansion resums,
by rewriting a perturbative series as a ratio of two polynomials.
Compared to the original expression, the PA introduces poles in
the coupling plane in a similar fashion as expected from renormalons.
This may explain why PAs seem to be able to approximately resum the
perturbative series, a finding corroborated e.g. by the
observation~\cite{ref:PADE} that measurements of the strong coupling from
global event shape variables become much more consistent when the second
order prediction is replaced by its $[1/1]$-PA.

\subsection*{Heavy Flavour Thresholds}
%=====================================
Another potentially very precise \as-measurement is derived from the
threshold behaviour of the $b\overline{b}$-production cross section in
\ee-annihilation. QCD sum rules for this process are dominated by the
non-relativistic threshold behaviour and allow to extract simultaneously
\as\/ and the $b$-quark mass. A determination of the strong
coupling~\cite{ref:BBTHR} at the BLM-optimized~\cite{ref:BLM}
renormalization scale $\mu=3$~GeV yields $\asmz=0.110 \pm 0.001$.
Naively one would set the renormalization scale to the $b$-mass,
which corresponds to a shift~\cite{ref:PDG96} $\delta\asmz=+0.008$. 
Taking this as the theoretical error of the measurement one has
$\asmz=0.110\pm 0.008$\/ or $\asmu=0.217^{+0.036}_{-0.030}$\/ for 
$\mu=3$~GeV, respectively.

\subsection{Bound States of Heavy Quarks}
%========================================
The strong coupling constant has been determined from ratios of decay 
rates, which are described by perturbative QCD, and level splittings 
between different radial excitations of the bound system, which probe 
the QCD-potential between quark and antiquark.

\subsection*{Decays of Heavy Quarkonia}
%======================================
A precise \as-measurement is based on the ratio of the hadronic 
to the leptonic width of a heavy quark-antiquark pair. At Born-level this
ratio of annihilation rate into three gluons over the rate into a lepton
pair is proportional to ${\cal O}(\alpha_s^3/\alpha_{em}^2)$. Taking into 
account relativistic corrections proportional to the average $<v^2/c^2>$\/ 
of the quarks, the theoretical prediction can be written in the form
$$
    \frac{\Gamma(q\bar{q} \into hadrons)}{\Gamma(q\bar{q} \into \ee)} 
  = R_{pert}
    \left(1 + D\left\langle \frac{v^2}{c^2} \right\rangle \right).
$$
The perturbative prediction $R_{pert}$ is known to NLO, $D$\/ is a 
free parameter for the size of the non-perturbative relativistic 
corrections. Assuming $D$\/ to be a universal constant, it has been 
extracted~\cite{ref:QQMK} together with \as\/ in a combined analysis of 
$\Upsilon$\/ and J/$\Psi$\/ decays. Theoretical uncertainties in 
$R_{pert}$\/ were studied by introducing ad hoc NNLO terms, by varying 
the renormalization scale and by Pad\'e-like rewriting terms of the type 
$(1+B\as)$\/ as $1/(1-B\as)$. The result at a scale $\mu=10$~GeV is 
$\asmu=0.167^{+0.015}_{-0.011}$, dominated by the theoretical uncertainties.

The CLEO collaboration~\cite{ref:QQCLEO} has extracted a 
measurement of the strong coupling from the ratio 
$\Gamma(\Upsilon \into gg\gamma)/\Gamma(\Upsilon \into ggg)$,
which to leading order is proportional to $\aem/\as$. The result at the
scale of the Upsilon mass $\mu=9.7$~GeV is $\asmu=0.163 \pm 0.009(\expt) 
\pm 0.010(\theo)$. The experimental error is dominated by systematic
uncertainties in the photon background of purely hadronic Upsilon decays
and roughly equal to the renormalization scale error.
The conservative average of both measurements taken at the scale 
$\mu=9.7$\/ is $\asmu=0.166 \pm 0.013$, which evolves to
$\asmz =0.112 \pm 0.006$.

\subsection*{Lattice Calculations}
%=================================
Very precise determinations of \as\/ were performed in the analysis 
of level splittings between the S- and the P-states of the
$\Upsilon$-system by means of lattice calculations.\cite{ref:QQLGT1} 
The current level of precision is a result of reduced lattice spacing 
errors, a better understanding of the conversion from the lattice to the 
\MSbar\/ coupling measured elsewhere and the introduction of fermion loops 
on the lattice. Calculations exist with $n_f=0$\/ and $n_f=2$\/ dynamic 
fermions, which give only marginally different results and thus allow a safe
extrapolation to the physical case of $n_f=3$\/ light flavours. The most
recent results~\cite{ref:QQLGT2} are $\asmz=0.118 \pm 0.003$~(NRQCD)
and $\asmz = 0.116 \pm 0.003$~(FNAL). The conservative average is
$\asmz=0.117 \pm 0.003$, which at the typical scale of the calculations
$\mu=8.2$~GeV corresponds to $\asmu=0.184 \pm 0.008$.

\subsection{Scaling Violations}
%==============================
Scaling violations are observed in Deep Inelastic Scattering (DIS)
processes in the structure functions of the nucleon, and in 
\ee-annihilation into hadrons in the fragmentation functions of the
primary partons, i.e. in reactions with space- and time-like momentum
transfers. In both cases perturbative QCD predicts a softening with
increasing $Q^2$, $d\ln F/d\ln Q^2\propto\as(Q)$\/ as described by the
Dokshitzer-Gribov-Lipatov-Altarelli-Parisi (DGLAP) evolution 
equations.\cite{ref:DGLAP} The softening of structure functions comes 
about because higher momentum transfers resolve more partons from vacuum
fluctuations in the nucleon. For fragmentation functions the growing
phase space permits additional gluon radiation and thus enhanced particle 
multiplicities in jets. The theory~\cite{ref:SCALTH} is known to
next-to-leading order. In a determination of \as\/ perturbative and 
non-perturbative effects can be disentangled through their energy 
dependence. Here DIS processes are favoured because non-perturbative 
effects decrease rapidly with $1/Q^2$. In \ee-annihilation they 
decrease only proportional to $1/Q$, which renders measurements of \as\/ 
from fragmentation functions less precise than the ones from DIS, despite 
the fact that they are performed at larger momentum transfers.

\subsection*{Scaling Violations in Structure Functions}
%======================================================

Measurements of the strong coupling constant were performed in DIS
with neutrino beams and charged leptons on targets of heavy and light
nuclei. The most recent results for \LambdaDIS\/ from various
experiments~\cite{ref:DIS1,ref:DIS2,ref:DIS3,ref:DIS4} are collected in 
\tab{DIS} and displayed in \fig{DIS}. All numbers are in good agreement, 
although there is a slight trend towards larger values for more recent
measurements. The two most precise results are from a combined analysis
of SLAC-BCDMS measurements~\cite{ref:DIS2}, $\LambdaDIS=263\pm 42$~MeV,
and a recent re-analysis of CCFR data~\cite{ref:DIS4} which gives
$\LambdaDIS=346 \pm 58$~MeV. For the latter the logarithmic derivatives
of the structure functions together with the NLO QCD-fit are shown in
\fig{CCFR}. The value for \LambdaDIS\/ moved up by 136~MeV compared to
a previous publication~\cite{ref:GLS4} based on the same data, mainly
as a consequence of an improved energy calibration for the detector.
The weighted average based on the measurements published after 1990
yields $\LambdaDIS = 287 \pm 31$~MeV. With a common theoretical
uncertainty~\cite{ref:DIS2} of $\delta\asmz=0.004$\/ this corresponds to
$\asmz=0.116 \pm 0.005$, or equivalently $\asmu = 0.200 \pm 0.016$\/
for an average scale $\mu = 5.4$~GeV. The total error from this NLO analysis 
is still dominated by theoretical uncertainties. First results from 
NNLO-analyses~\cite{ref:DIS5} indicate that the value of \as\/ is quite 
stable when going to higher orders, i.e. the theoretical error can be 
expected to decrease. 

\ren{\arraystretch}{1.25}
\begin{table}[t]
\begin{center}
\small
\begin{tabular}{|l|l|c|l|}
\hline
 \multicolumn{2}{|c|}{Measurement} &
 \multicolumn{1}{|c|}{$\mu$/GeV}    &
 \multicolumn{1}{|c|}{\LambdaDIS/MeV} \\
\hline
 CDHS       &  $\nu$Fe               & 4.5 & \errax{250}{150}{100}  \\
 CHARM      &  $\nu$Ca               & 3.9 & \errsx{310}{157}       \\
 BEBC       &  $\nu$Ne               & 4.0 & \errax{100}{110}{85}   \\
 BCDMS      &  $\mu$C                & 8.4 & \errsx{230}{63}        \\
\hline 
 EMC        &  $\mu$H$_2$            & 4.7 & \errax{211}{117}{108}  \\
 SLAC-BCDMS &  $l$H$_2$,$l$D$_2$     & 7.1 & \errsx{263}{42}        \\
 CHDSW      &  $\nu$Fe               & 6.7 & \errsx{300}{100}       \\
 NMC        &  $\mu$H$_2$,$\mu$D$_2$ & 2.6 & \errax{306}{199}{188}  \\
 CCFR       &  $\nu$Fe               & 8.2 & \errsx{346}{58}        \\
\hline
 \multicolumn{2}{|c|}{Average}       & 5.4 & \errsx{287}{31}        \\
\hline
\end{tabular}
\normalsize
\end{center}
\caption{\sf Measurements of the QCD scale from DIS in chronological
             order. The errors are the purely experimental uncertainties.
             The average is the weighted average from the lower part
             of the table.}
\label{tab:DIS}
\end{table}
\ren{\arraystretch}{1.0}

\begin{figure}[h]
\begin{center}
   \fplace{0.45}{0.0cm}{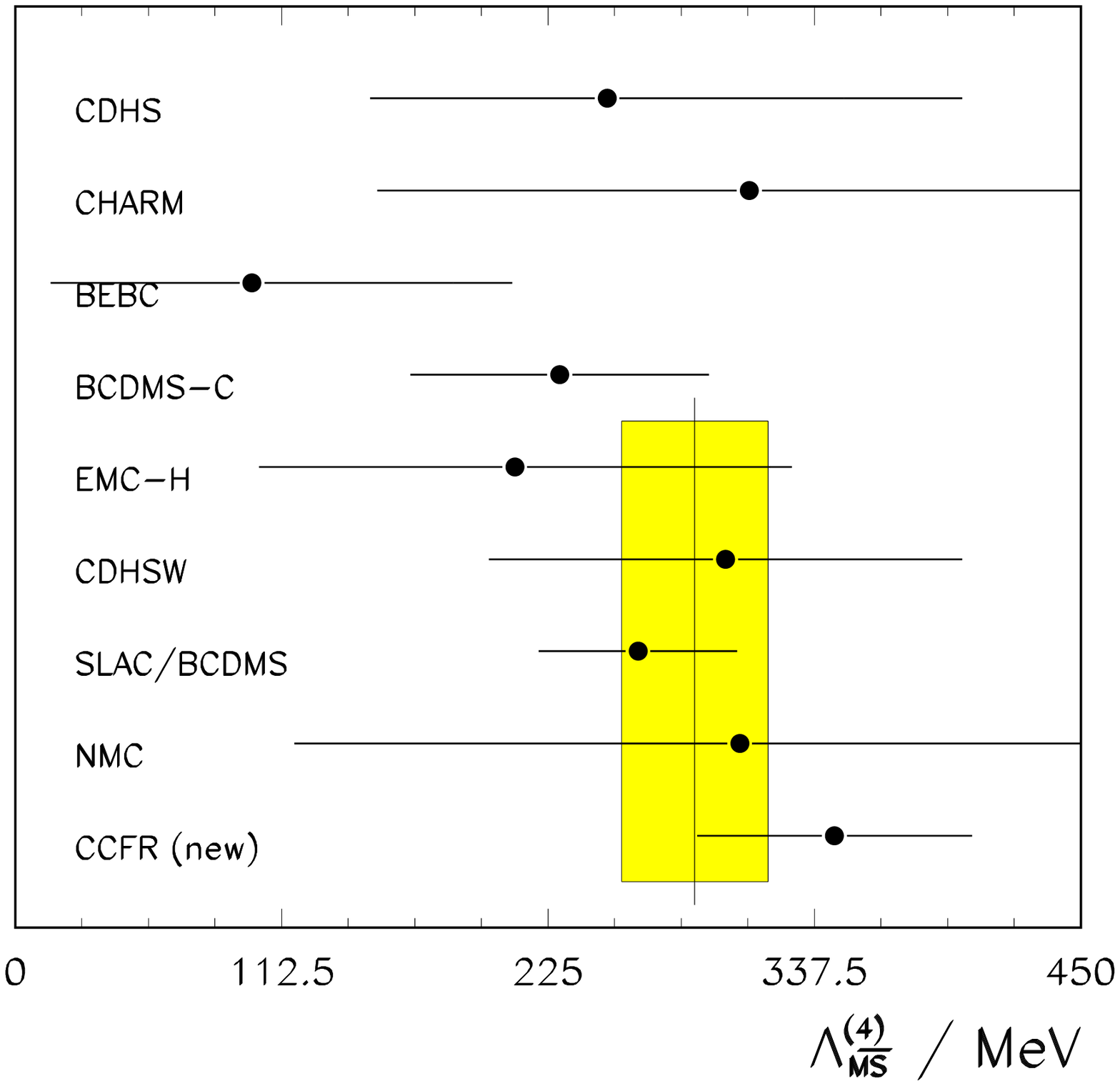}
   \cplace{0.45}{3.0cm}{fig:DIS}
          {\sf Graphical representation of the results given in 
               table~\ref{tab:DIS}. The vertical line marks the average 
               \LambdaDIS, the shaded area its uncertainty.}  
\end{center}
\end{figure}

\begin{figure}[t]
\begin{center}
   \fplace{0.475}{0.cm}{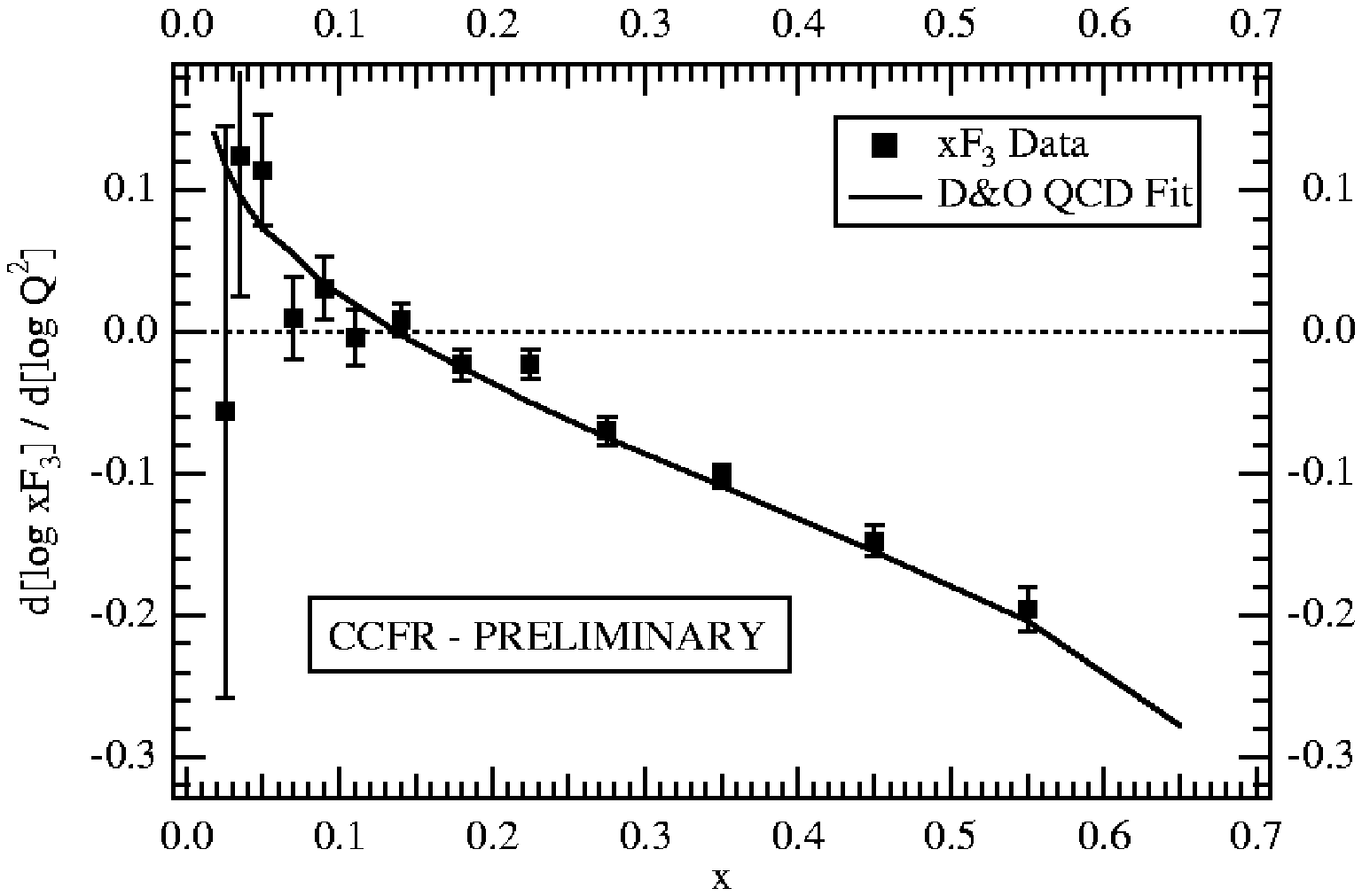} 
   \fplace{0.475}{0.cm}{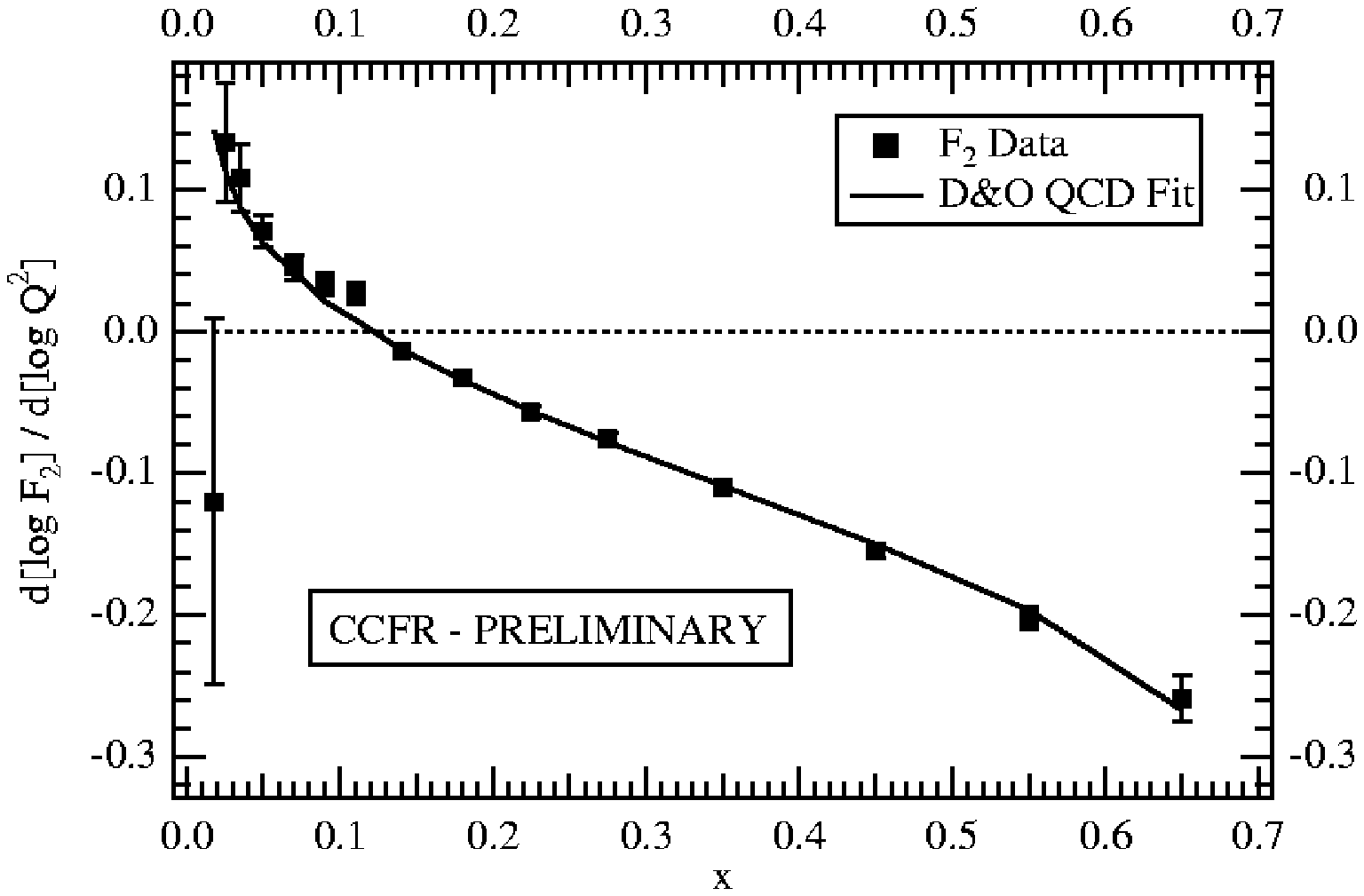} 
   \cplace{0.750}{0.cm}{fig:CCFR}
          {\sf Logarithmic derivatives of the structure functions $F_2$\/ 
               and $xF_3$\/ as measured by the CCFR collaboration. The 
               solid line is the result of a NLO QCD-fit.}
\end{center}
\end{figure}

\subsection*{Scaling Violations in Fragmentation Functions}
%========================================================== 
The full NLO theoretical framework has also been used in two
determinations of the strong coupling constant from scaling violations in
frag\-men\-tation functions.\cite{ref:SCALV1,ref:SCALV2} Both are based on 
inclusive and $uds$, $c$, $b$ and gluon-jet enriched data samples from 
hadronic Z
decays, combined with measurements from lower centre-of-mass energies down
to $\sqrt{s}=22$~GeV or $\sqrt{s}=14$~GeV, respectively. The strong
coupling constant was determined together with parametrizations for the
fragmentation functions of the different parton types and a power-law
correction describing non-perturbative effects. The latter was found to
be small for the energy range under consideration. The results of these
model independent analyses are~\cite{ref:SCALV1} $\asmz=0.126 \pm 0.009$\/
and~\cite{ref:SCALV2} $\asmz=0.121 \pm 0.012$, where the error is
dominated by the QCD scale uncertainty. The larger error from the
second analysis reflects a larger range for scale variations. The
conservative average of both results is $\asmz = 0.124 \pm 0.010$, which
for a central scale $\mu=36$~GeV corresponds to $\asmu=0.146 \pm 0.014$.

Assuming that non-perturbative effects and differences in the
fragmentation between quarks of different flavours and gluons are
described correctly by Monte-Carlo models, one can also exploit the
LUND matrix element model as the theoretical basis for a measurement
of \as\/ from scaling violations. A result~\cite{ref:SCALV3} obtained
by the DELPHI collaboration is consistent with the other determinations.
Finally, it is worth noting that scale breaking effects in fragmentation
functions can also be studied in DIS, using the momentum distribution of
particles in the current jet.\cite{ref:SCALV4} DIS thus offers the
possibility to measure \as\/ from scale breaking effects both in
structure functions and in fragmentation functions.

\subsection{Results from Hadron Colliders}
%=========================================
Prompt photon production in hard parton-parton scattering is a
Compton-like process of ${\cal O}(\as\aem)$. In the difference 
$\sigma(\pbp\into\gamma X)-\sigma(pp\into\gamma X)$\/ the
sea quark and gluon structure functions of the nucleon cancel, i.e. only
the well known valence quark distribution is needed as external input
for an \as-determination. The UA6 collaboration~\cite{ref:UA6} finds
$\LambdaDIS = 235 \pm 106(\expt)^{+146}_{-9}(\theo)$\/~MeV. Translated
into \as\/ at the typical scale $\mu=4$~GeV of the measurement one finds
$\asmu = 0.206^{+0.042}_{-0.033}$, which corresponds to
$\asmz=0.112^{+0.012}_{-0.010}$. 

Heavy quarks not present in the initial state can be produced by 
quark-antiquark annihilation or gluon-gluon fusion processes, which 
to leading order are of \Oas{2}. The theory is developed 
to NLO. Experimentally those reactions can be tagged by the decay 
characteristics of the heavy hadrons. The most precise 
measurement~\cite{ref:UA1} from an analysis of $b\overline{b}+$jets 
production in $p\bar{p}$\/ collisions is $\asmu=0.138^{+0.028}_{-0.019}$,
determined at a scale $\mu=20$~GeV. Evolved to the Z mass one finds
$\asmz=0.109^{+0.016}_{-0.012}$.

The strong coupling constant has also been determined from QCD-radiative
corrections to W production at hadron colliders, based on the cross section 
ratio $R_W=\sigma(W+1\,\mbox{jet})/\sigma(W+0\,\mbox{jets})$. This ratio is 
sensitive to the product $\as F(x)$\/ of strong coupling constant and parton 
densities in the nucleon. The perturbative QCD-correction is known to NLO. 
Taking the nucleon structure functions from low energy data, \as\/ enters 
both in the matrix element for jet-production in W decays and in the 
evolution of the structure functions to collider energies. A measurement 
$\asmw = 0.123 \pm 0.025$, or equivalently $\asmz=0.121\pm 0.024$, has been 
obtained by the UA2-collaboration~\cite{ref:UA2} at a centre-of-mass energy 
$\sqrt{s}=630$~GeV, with an error dominated by experimental uncertainties. 
The ratio $R_W$\/ has also been determined at the Tevatron~\cite{ref:D0}, 
but it turns out to be rather insensitive to \as\/ when parton densities
from low energy measurements are employed in the analysis. The reason is 
that for larger values of \as\/ the enhanced softening of the structure 
functions in the evolution almost compensates the increased jet production 
rates from the matrix element. In other words, the determination of the 
strong coupling from W-decays at the Tevatron requires parton densities 
measured at the same energy.

Another interesting prospect is the determination of the strong coupling 
constant from the inclusive jet cross section at hadron colliders.
Here an extraction of $\alpha_s(E_T)$\/ with high precision seems
feasible.\cite{ref:JETET}

\subsection{Global Event Shape Variables}
%========================================
To leading order the fraction of three-jet events in \ee-annihilation is 
proportional to \as. Global event shape variables which are sensitive to the 
topology of multijet events exploit this fact for a measurement of the strong 
coupling constant. In order to be defined in perturbative QCD they must be 
constructed such, that they are insensitive to soft gluon radiation and that 
they remain unchanged if any of the final state particles splits into two 
collinear ones. An example is the variable thrust~\cite{ref:DEFT}
$T=\max_{\vec{n}}(\sum_p| \vec{p}\cdot\vec{n}|)/(\sum_p |\vec{p}|)$,
which measures the collimation of the momentum flow in an event.
An ideal two-jet event without final state gluon radiation has $T=1$.
Gluon radiation decreases the value of $T$\/ until one finds $T=1/2$\/
for a perfectly isotropic event.

Another example is $y_3$, defined by means of a jet clustering
algorithm where initially each particle is considered its own jet. Then
those two jets which are closest in phase space are combined by adding
their 4-momenta. Iterating the procedure, $y_3$\/ is defined as that 
distance where the event makes the transition from three to two jets. 
Common measures for the distance between two jets $i$\/ and $j$\/ are the 
Durham-metric~\cite{ref:DURHAM} 
$y^D_{ij}=2\min(E_i^2,E_j^2)(1-\cos\theta_{ij})/s$\/ or the 
Jade-metric~\cite{ref:JADE} $y^J_{ij}=2(E_i E_j)(1-\cos\theta_{ij})/s$.
In both cases  $\theta_{ij}$\/ is the opening angle between the jets and
$s$\/ denotes the total invariant mass of the hadronic system. 

The theoretical prediction for all global event shape variables is known
to NLO, based on a numerical integration~\cite{ref:EVENT}
of the ERT-matrix elements.\cite{ref:ERT} For some variables also
leading-logarithmic and next-to-leading logarithmic corrections, which
dominate the cross section at high thrust, have been resummed to all
orders.\cite{ref:RESUM} In the latter case an improved theoretical 
prediction is obtained by combining the resummed prediction with the second 
order matrix elements, which is exact to \Oas{2}\/ over the whole phase 
space and contains the dominant terms for two-jet like configurations to 
all orders. There is a certain freedom in performing the 
matching,\cite{ref:MATCHING} which gives rise to differences at \Oas{3}. 
It thus can be employed as an alternative to the variation of the 
renormalization scale to probe the sensitivity of an \as\/ measurement to 
unknown higher order perturbative corrections.

The non-perturbative transition from partons to hadrons gives rise to 
power-law corrections, which in contrast to the case of inclusive variables 
cannot be handled in the framework of the OPE and thus usually are estimated 
by means of Monte Carlo models.\cite{ref:GENERATORS} This dependence on 
phenomenological models introduces an additional ``hadronization 
uncertainty'' into an \as-measurement. Only recently analytic 
calculations~\cite{ref:POWEE} became available, based on the study of 
the long-distance behaviour of leading order matrix elements, 
which in some respect can be understood as generalizations 
of the OPE. Identifying different classes of power-law corrections, they 
introduce a small number of universal parameters, e.g. effective values for 
integrals over powers of the strong coupling at low energies, which relate
non-perturbative corrections for different event shape variables.

The formalism has been worked out for mean values of some global event
shape variables. From a measurement of the energy dependence of those mean
values, a model independent determination of the strong coupling is
obtained by fitting \as\/ together with the non-perturbative power law 
corrections. The result of a first analysis of this kind is shown 
in \fig{ASDELPHI}. 

\begin{figure}[htb]
\begin{center}
   \fplace{0.55}{0.0cm}{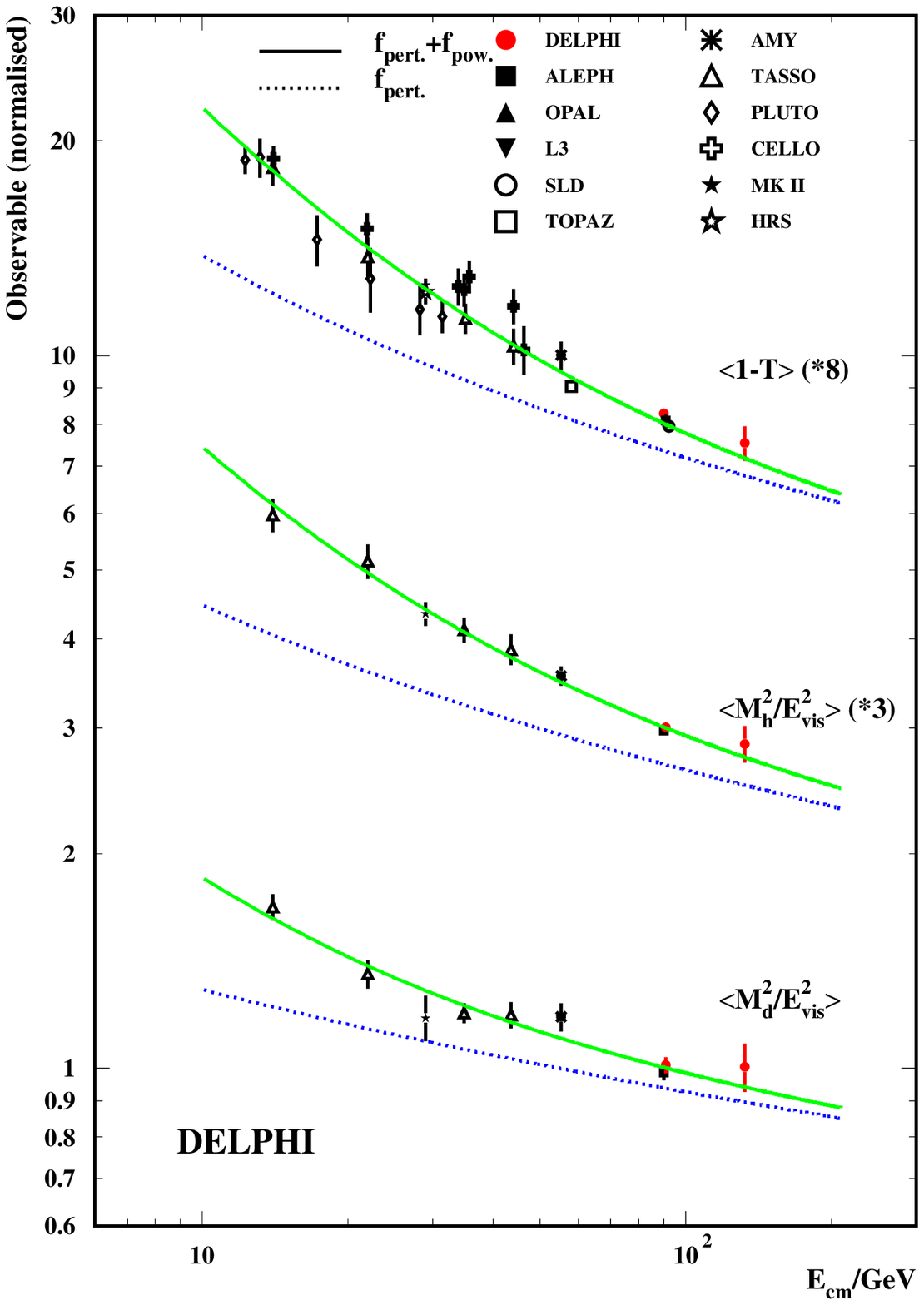} \mm{5}
   \cplace{0.35}{8.8cm}{fig:ASDELPHI}
          {\sf Measurement of \as\/ based on mean values of global
            event shape variables measured between $\sqrt{s}=14$~GeV
            and $\sqrt{s}=133$~GeV.}
\end{center}
\end{figure}

Measurements of \as\/ based on global event shape 
variables~\cite{ref:EVSH1,ref:EVSH2,ref:EVSH3,ref:EVSH4,ref:EVSH5,ref:EVSH6}
from centre-of-mass energies between $\sqrt{s}=10.53$~GeV and
$\sqrt{s}=133$~GeV are listed in \tab{ASEVSH}. Wherever available, 
the numbers are single experiment averages over several variables. 
Combining the partial averages from energies in the range
$10.53\GeV\leq\sqrt{s}\leq 133\GeV$\/ into one global mean value 
finally yields $\asmz=0.121\pm 0.005$.

\begin{table}[htb]
\begin{center}
\small
\begin{tabular}{|l|l|c|c|}
\hline
                & Theory  & $\mu$/GeV & $\alpha_s(\mu)$     \\
\hline
  CLEO          & NLO     & 10.53& $0.164 \pm 0.015$ \\
\hline
  TPC/$2\gamma$ & NLO+NLL & 29.0 & $0.160 \pm 0.012$ \\
\hline
  MAC           & NLO     & 34.0 & $0.130 \pm 0.032$ \\
  MARKII        & NLO     & 34.0 & $0.153 \pm 0.032$ \\
  PLUTO         & NLO     & 34.0 & $0.108 \pm 0.038$ \\
  TASSO         & NLO     & 34.0 & $0.149 \pm 0.026$ \\
  MARK-J        & NLO     & 34.0 & $0.126 \pm 0.013$ \\
  CELLO         & NLO     & 34.0 & $0.144 \pm 0.026$ \\
  JADE          & NLO     & 34.0 & $0.162 \pm 0.043$ \\
\hline
 \multicolumn{2}{|c|}{Average} & 34.0 & $0.134 \pm 0.019$ \\  
\hline
  AMY           & NLLJET  & 58.0 & $0.130 \pm 0.006$ \\
  TOPAZ         & NLO+NLL & 58.0 & $0.132 \pm 0.008$ \\
  TOPAZ         & NLO+NLL & 58.0 & $0.139 \pm 0.008$ \\
  VENUS         & NLLJET  & 58.0 & $0.129 \pm 0.006$ \\
\hline
 \multicolumn{2}{|c|}{Average} & 58.0 & $0.132 \pm 0.006$ \\  
\hline
  ALEPH         & NLO+NLL & 91.2 & $0.125 \pm 0.005$ \\
  DELPHI        & NLO+NLL & 91.2 & $0.123 \pm 0.006$ \\
  L3            & NLO+NLL & 91.2 & $0.124 \pm 0.007$ \\  
  OPAL          & NLO+NLL & 91.2 & $0.120 \pm 0.006$ \\
  SLD           & NLO+NLL & 91.2 & $0.120 \pm 0.008$ \\
\hline
 \multicolumn{2}{|c|}{Average} & 91.2 & $0.123 \pm 0.006$ \\    
\hline
  ALEPH         & NLO+NLL & 133. & $0.119 \pm 0.008$ \\
  DELPHI        & NLO+NP  & 133. & $0.119 \pm 0.009$ \\
  L3            & NLO+NLL & 133. & $0.107 \pm 0.010$ \\  
  OPAL          & NLO+NLL & 133. & $0.110 \pm 0.008$ \\
\hline
 \multicolumn{2}{|c|}{Average} & 133. & $0.114 \pm 0.007$ \\  
\hline
\end{tabular}
\normalsize
\end{center}
\caption[]{\sf
         Measurements of the strong coupling constant from global
         event shape variables. The errors are the total errors
         of the individual measurements, which are dominated 
         by the theoretical uncertainties. The averages were formed
         as described in the introduction. The results from AMY and 
         VENUS are based on theoretical predictions calculated by the 
         NLLJET Monte Carlo. The DELPHI-analysis at $\sqrt{s}=133$~GeV
         uses the second order prediction from perturbative QCD 
         together with an analytical ansatz for non-perturbative effects.}
\label{tab:ASEVSH}
\end{table}

\subsection{Jets from processes with variable $Q^2$}
%===================================================
One contribution to multijet production in $ep$-collisions is gluon radiation 
from a quark scattered off a virtual photon with large $Q^2$. Quark
and gluon emerge as two jets in addition to the jet from the proton remnant.
The production rate $R_{2+1}$\/ of those (2+1)-jet final states is known
to NLO for jets defined by the JADE algorithm. Tagging the scattered electron 
allows to control the $Q^2$\/ of the process and thus to establish the running
of the strong coupling constant within a single experiment. First measurements
have been published by the H1 and ZEUS collaborations.\cite{ref:ASEP} The 
conservative average of both results is $\asmz =0.119 \pm 0.013$, which 
corresponds to $\asmu=0.156 \pm 0.022$\/ for a central scale  
$\mu=19.6$~GeV of the two measurements. The error is dominated by the 
theoretical uncertainties. The published data still suffer from too low 
statistics for a convincing demonstration of the running of \as, but 
improved results can be expected within the near future.\cite{ref:ASEP2}

Using radiative Z-decays, $Z\into q\overline{q}\gamma$, the running
of the strong coupling can also be seen in \ee-annihilation by a
single experiment. A first measurement with still rather low statistical
precision has been presented by the L3-collaboration.\cite{ref:ASEE}
The strong coupling extracted at six different scales below the Z-mass
was found to be consistent with the running as expected by QCD. The
combined result is $\asmz=0.119 \pm 0.007$, which corresponds to 
$\asmu=0.131 \pm 0.008$\/ for an average scale $\mu=50.8$~GeV.

\section{A Global Average for \AS}
%=================================
\ren\arraystretch{1.1}
\begin{table}[tb]
\begin{center}
\small
\begin{tabular}{|l|c|l|l|}
\hline
   \multicolumn{1}{|c|}{Measurement}
 & \multicolumn{1}{|c|}{$\mu$/GeV}  
 & \multicolumn{1}{|c|}{\asmu}
 & \multicolumn{1}{|c|}{\asmz}  \\
\hline
 BjSR              & 1.732 & \erra{0.320}{35}{55} & \erra{0.118}{05}{07} \\
 GLSR              & 1.732 & \erra{0.260}{41}{47} & \erra{0.110}{06}{09} \\
 \Rtau             & 1.777 & \errs{0.347}{37}     & \errs{0.122}{05}     \\
 \bb\/ threshold   & 3.    & \erra{0.217}{36}{30} & \errs{0.110}{08}     \\ 
 prompt $\gamma$   & 4.0   & \erra{0.206}{42}{33} & \erra{0.112}{12}{10} \\ 
 DIS               & 5.4   & \erra{0.200}{16}{15} & \errs{0.116}{05}     \\
 LGT               & 8.2   & \errs{0.184}{08}     & \errs{0.117}{03}     \\
 \cc, \bb\/ decays & 9.7   & \errs{0.166}{13}     & \errs{0.112}{06}     \\
 \Rgamma           & 18.0  & \errs{0.175}{23}     & \erra{0.128}{12}{13} \\
 $ep\into$Jets     & 19.6  & \errs{0.156}{22}     & \errs{0.119}{13}     \\ 
 \pp\into\bb+Jets  & 20.0  & \erra{0.138}{28}{19} & \erra{0.109}{16}{12} \\
 \ee fragm.        & 36.0  & \errs{0.146}{14}     & \errs{0.124}{10}     \\
 \pp\into W+Jets   & 80.2  & \errs{0.123}{25}     & \errs{0.121}{24}     \\
 SM constraints    & 91.2  & \errs{0.120}{03}     & \errs{0.120}{03}     \\
 event shapes      & $\leq$ 133 &                 & \errs{0.121}{05}     \\
\hline
\end{tabular}
\normalsize
\caption{\sf
         Compilation of measurements of the strong coupling constant.
         The total errors in the last two digits as given in parenthesis
         in most cases are dominated by the theoretical uncertainties.}
\label{tab:ASGLOB}
\end{center}
\end{table}
\ren\arraystretch{1.0}

A compilation of the results discussed in the preceeding section is given in 
\tab{ASGLOB}. The \as-values from the different types of measurements evolved 
to the scale of the Z-mass are displayed in \fig{ASGLOB}. Over an energy 
range from $\mu=1.73$~GeV to $\mu=133$~GeV all results appear to be consistent 
with one common mean for \asmz. The weighted average, assuming uncorrelated 
errors, is $\asmz=0.1177 \pm 0.0015$\/ with $\chi^2/{\rm ndf}=6.7/14$. The 
conservative average is  $\asmz=0.1177\pm 0.0037$, where this error is biased 
towards large values through the inclusion of some less precise results. 
Restricting the average to measurements with error $\delta\asmz < 0.008$, 
one finally obtains $\asmz=0.1183 \pm 0.0032$.

The value of the strong coupling constant at the physical scale of the 
respective measurement is shown in \fig{ASDRUN}. The running in agreement 
with the QCD prediction is evident. Note that many different types of 
reactions with different numbers of active quark flavours are consistent,
in agreement with the expectation that \as\/ is flavour independent.

\begin{figure}[htb]
\begin{center}
   \fplace{0.80}{0.cm}{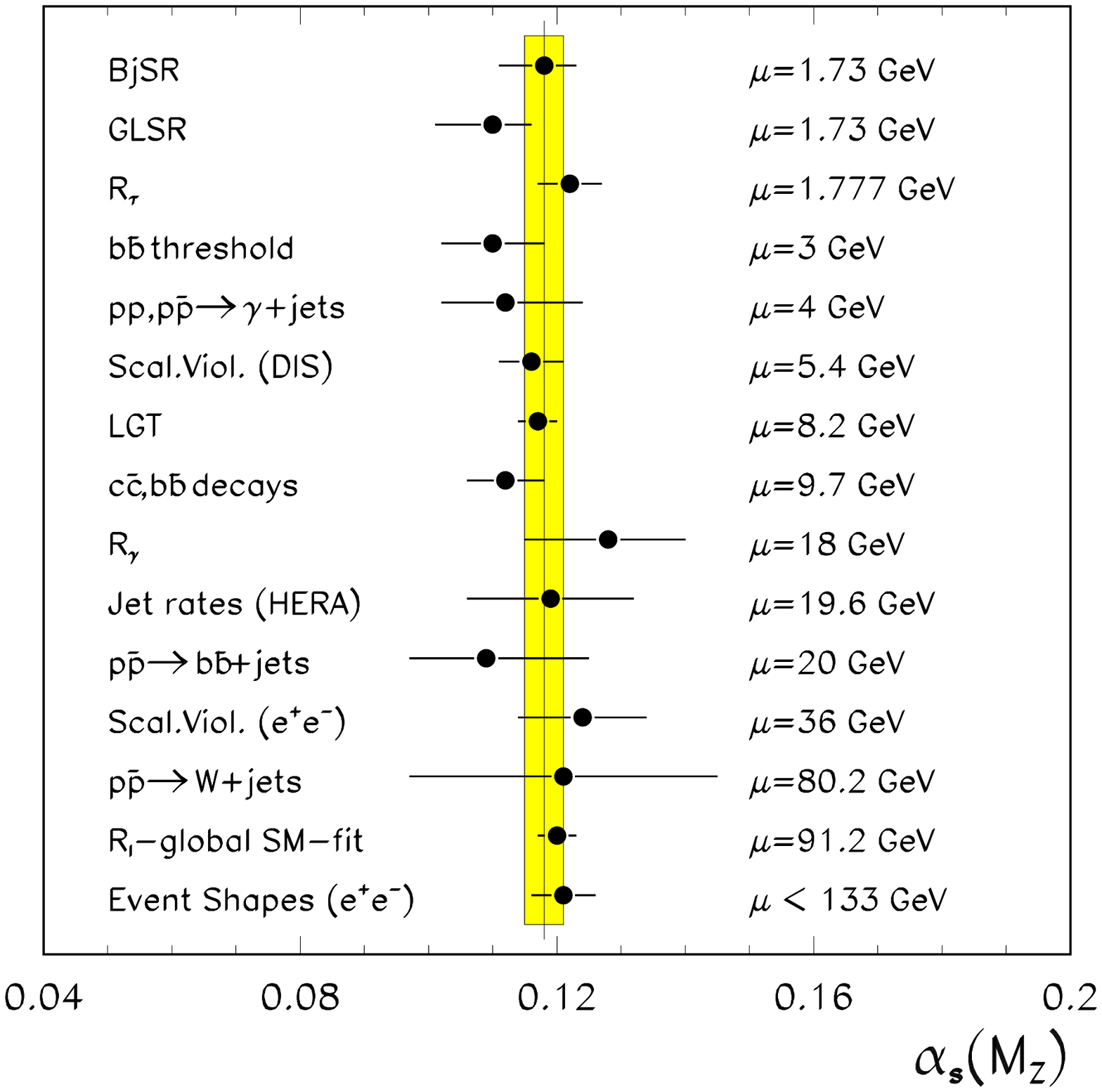}
   \cplace{0.90}{0.cm}{fig:ASGLOB}
          {\sf Measurements of the strong coupling constant from different 
               sources evolved to the scale of the Z-mass.}
\end{center}
\end{figure}

\begin{figure}[ht]
\begin{center}
   \fplace{0.99}{0.cm}{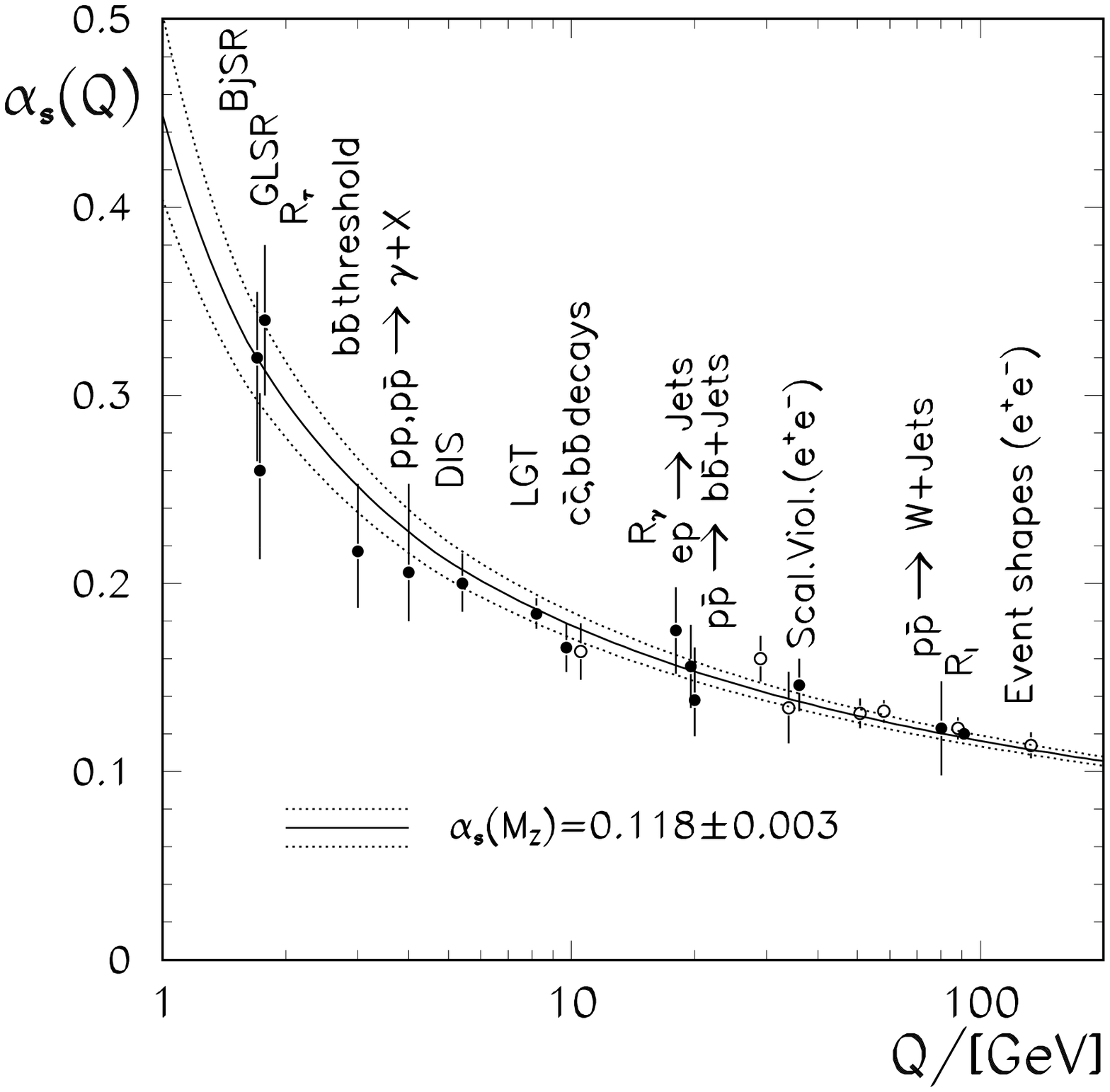} \\
   \cplace{0.99}{0.cm}{fig:ASDRUN}
          {\sf Running of the strong coupling constant established by  
               various types of measurements at different scales, compared 
               to the QCD prediction for $\asmz=0.118 \pm 0.003$. The open 
               dots are results based on global event shape variables.}
\end{center}
\end{figure}

\section{Summary}
%================
In the past year significant progress has been made in the determination
of the strong coupling constant by an improved understanding of power
law corrections and perturbative higher orders on the theoretical side, and
by precise new measurements from DIS, lattice gauge theories and standard
model fits. All results are perfectly compatible with a common mean. Based 
on data covering the energy range $1\GeV < \mu < 133\GeV$, an average value
$\asmz=0.118 \pm 0.003$\/ is obtained. The error takes into account the
possibility of positive correlations between the measurements. From the
RMS-scatter of the individual results an even smaller error would be 
derived.

\small
\subsection*{Acknowledgments}
%=========================
I would like to thank the organizers of the ICHEP 1996 conference for the 
pleasant atmosphere and good working conditions in Warsaw, the Max-Planck 
Gesellschaft for financial support, and all my colleagues, in particular 
Debbie Harris and G\"unther Quast, who helped to provide latest results. 
My sincere apologies go to everybody whose contribution to the wide field 
addressed in this paper I failed to mention.

\end{document}